\documentclass{article}
\usepackage{authblk}

\expandafter\let\csname equation*\endcsname\relax
\expandafter\let\csname endequation*\endcsname\relax

\usepackage{mathtools}
\usepackage{graphicx}
\usepackage{float}
\usepackage[labelfont=bf]{caption}
\usepackage{bm}
\usepackage{xcolor}
\usepackage[normalem]{ulem}
\usepackage{braket}
\usepackage{cite}
\usepackage{booktabs} 
\usepackage{hyperref}
\usepackage{amssymb}
\usepackage{gensymb}
\usepackage{siunitx}
\usepackage[colorinlistoftodos]{todonotes}
\usepackage{setspace}
\usepackage{etoolbox}
\pretocmd{\abstract}{\newpage}{}{}
\usepackage{tabularx}

\begin{document}
\title{Large scale FRET simulations reveal the control parameters of phycobilisome light harvesting complexes}

\author{Emma Joy Dodson\textsuperscript{1}, Nicholas Werren\textsuperscript{2}, Yossi Paltiel\textsuperscript{3}, Erik Gauger\textsuperscript{2,*},
Nir Keren\textsuperscript{1,*}}

\affil{\textsuperscript{1}Department of Plant and Environmental Science, The Alexander Silberman Institute of Life Sciences, The Hebrew University in Jerusalem, Jerusalem, Israel}
\affil{\textsuperscript{2}SUPA, Institute of Photonics and Quantum Sciences, Heriot-Watt University, Edinburgh, EH14 4AS, United Kingdom
}
\affil{\textsuperscript{3}Department of Applied Physics, The Hebrew University in Jerusalem, Jerusalem, Israel}

\affil{\textsuperscript{*}Corresponding authors: e.gauger@hw.ac.uk, nir.ke@mail.huji.ac.il}

\date{}

\maketitle

\doublespacing
\begin{abstract}
Phycobilisomes (PBS) are massive structures that absorb and transfer light energy to photochemical reaction centers. 
Among the range of light harvesting systems, PBS are considered to be excellent solutions for absorption cross-sections but relatively inefficient energy transferring systems.
This is due to the combination of a large number of chromophores with intermediate coupling distances.
Nevertheless, PBS systems persisted from the origin of oxygenic photosynthesis to present day cyanobacteria and red algae, organisms that account for approximately half of the primary productivity in the ocean. 
In this study we modeled energy transfer through subsets of PBS structures, using a comprehensive dynamic Hamiltonian model.
Our approach was applied, initially, to pairs of phycobilin hexamers and then extended to short rods.
By manipulating the distances and angles between the structures we could probe the dynamics of exciton transfer.
These simulations suggest that the PBS chromophore network enhances energy distribution over the entire PBS structure – both horizontally and vertically to the rod axis.
Furthermore, energy transfer was found to be relatively immune to the effects of distances or rotations, within the range of intermediate coupling distances. Therefore, we suggest that the PBS provides unique advantages and flexibility to aquatic photosynthesis. 

\end{abstract}

\section{Introduction}

In the photosynthetic process, light-harvesting pigment-protein complexes absorb light energy and transfer it to photosystems, where photochemical reactions occur \cite{Glazer85}.
While the structure of the photosystems is highly conserved in evolution, light-harvesting is done by a broad and diverse array of complexes \cite{Hohmann11}.
One of the most prevalent light-harvesting systems is the phycobilisome (PBS), present in cyanobacteria and red algae.
It is composed of soluble proteins anchored to the photosynthetic membrane surface and chromophores -- tetrapyrrole molecules that covalently attach to conserved sites on the proteins \cite{MacColl98}.
The spectral properties of the chromophores are determined by their chemical nature and by their interaction with proteins \cite{Adir20}.
The basic building blocks of PBS structures are phycobilin chromophore binding proteins which assemble into hexamers (see Figure \ref{fig:mod_diagram} for hexamer visualization).
These hexamers either organize to form the core of the PBS antenna or stack into rod structures with the help of mostly unpigmented linker proteins \cite{Adir20}.
Antenna rods channel excitonic energy to the PBS core for transfer to the photochemical reaction centers embedded in the thylakoid membranes \cite{Grondelle94}.

The efficiency of energy transfer through light harvesting systems is a major determinant of the overall efficiency of the photosynthetic process.
Among light harvesting systems, the PBS are not considered to be very efficient.
The number of chromophores in a single antenna system can reach well above 1000 \cite{Ma20}, creating one of the largest known absorption cross-sections for the photosynthetic unit \cite{Mauzerall89}.
According to FRET random walk principles, the diffusion length of an exciton increases with the number of pigments in a photosynthetic unit \cite{Feron12}.
However, experimental results of energy transport in photosystems give a more nuanced picture wherein efficiency and yield are demonstrated to be constrained by the physical configuration of the system and its surroundings
\cite{Wientjes13}.
The chromophores of PBS are relatively far apart, generating a network of intermediately coupled chromophores.
Furthermore, PBS rods can reach four hexamers in length, while FRET random walk calculations demonstrate that an antenna rod extending beyond three hexamers in length will exceed the limit for maintaining energy-efficient transfer \cite{Chenu17}.
However, it is important to note that high efficiency should not be considered a goal in and of itself for photosynthesis in natural environments \cite{Keren18}.
Large dynamic range and robustness of energy transfer can be equally or even more important for the fitness of a photosynthetic organism.

Over the past few decades, different approaches have been applied to model and simulate excitation energy transfer (EET) of the PBS \cite{Figueroa12, Matamala07, Padyana06, Ren13, Debreczeny95}.
In these studies, models of protein hexamers and short rods were used to simulate energy flow with calculations based on F\"orster coupling between the transition dipoles of the interacting chromophores as well as their spectral properties.
While the authors of these studies were able to propose pathways for energy transfer in the system, the simulations were limited due to missing structural details of intact PBS systems, particularly the exact conformation of hexamer stacking along with the orientation of the rods relative to each other.
In recent years, intact PBS structures have been resolved. The first intact structures were from red algae \cite{Ma20, Zhang17}. These structures are massive, containing 14 rods and 1598 light-absorbing chromophore molecules. More recently, the Kerfeld lab resolved the structure of a cyanobacterial PBS containing four Orange Carotenoid Proteins (OCP) \cite{Dominguez22}.
These structures provide a good deal of here-to-unknown structural information and a basis for understanding changes in PBS efficiency, through energy transfer simulations \cite{Dominguez22}. However, these simulations did not calculate dynamic quantum effects. 

In this work, we present a framework for modelling energy transport with a quantum model.
By combining structural PBS data with a comprehensive Hamiltonian model, we capture exciton dynamics relevant to fundamental biological processes. 
The model described here utilises pairwise, distance-dependent dipole-dipole coupling (F\"orster) terms. Using this approach, we were able to efficiently run large-scale calculations to simulate the distribution and efficiency of PBS energy transfer which provide insight into the design principles that govern this immensely complex photosynthetic antenna system.

\section{Model} \label{sec:mod}

\subsection{Physical model} \label{sec:tback}

The approach outlined in the following section is similar to previous approaches used to model light-harvesting antennae and Iron-stress-induced-A proteins \cite{Dexter53, Grondelle94, Ishizaki09, Curutchet17, Brown19, Schoffman20}.
We reduce each chromophore in PBS down to an optical dipole located at the centre of mass from which we orientate the respective transition dipole vector.
These values are influenced by their position in the photosystem and the corresponding protein environment.
We therefore consider three different species of chromophore in this system: $\alpha$-84, $\beta$-84, and $\beta$-155.
The resulting description of the aforementioned hexamer structures in PBS is broken down into a collection of optical dipoles captured by the following Hamiltonian:
\begin{equation} \label{eq:Hs}
    \hat{H_{S}} = \sum^{N}_{j=1} \omega^{s}_{j} \hat{\sigma}_{j}^{z} + \sum^{N}_{j,k=1} J_{j,k}(\textbf{r}_{j,k}) \left(\hat{\sigma}_{j}^{+} \hat{\sigma}_{k}^{-} + \hat{\sigma}_{j}^{-} \hat{\sigma}_{k}^{+} \right)~,
\end{equation}
where the first term is the bare Hamiltonian of the chromophore sites, with species-dependent transition frequency $\omega^{s}_{j}$ at the $j$th site, and the second describes exciton hopping between the $j$th and the $k$th site, captured by the raising and lowering operators $\hat{\sigma}_{j}^{-}$ and $\hat{\sigma}_{k}^{-}$.
The strength of these hopping terms, the resonant F\"orster interactions, are given by $J_{j,k}(\textbf{r}_{j,k})$:
\begin{equation} \label{eq:forst}
    J_{j,k}(\textbf{r}_{j,k}) = \frac{1}{4\pi \epsilon_{0}} \bigg( \frac{\textbf{d}_{j} \cdot  \textbf{d}_{k}}{|\textbf{r}_{j,k}|^{3}} - \frac{3 (\textbf{r}_{j,k} \cdot \textbf{d}_{j})(\textbf{r}_{j,k} \cdot \textbf{d}_{k})}{|\textbf{r}_{j,k}|^{5}} \bigg)~,
\end{equation}
for dipoles of oscillator strengths $\textbf{d}_j$ and $\textbf{d}_k$ and separated by the distance $\textbf{r}_{j,k}$ \cite{Curutchet17}.
Being closely spaced relative to the wavelength of optical photons, the chromophores can be assumed to interact collectively with a shared optical bath \cite{Cao20};
this coupling is captured by the optical interaction Hamiltonian \cite{Breuer02Book},
\begin{equation} \label{eq:HIopt}
    \hat{H}_{I,opt} = \sum^{N}_{k=1}\textbf{d}_{k} \hat{\sigma}_{k}^{x} \otimes \sum_{p} f_{p} (\hat{a}_{p} + \hat{a}_{p}^{\dag})~,
\end{equation}
where $\hat{a}_{p}^{(\dag)}$ is the annihilation (creation) operator for the $p$th optical mode and $f_{p}$ is the coupling strength between the $p$th mode and the chromophores.
Furthermore, the dipoles are each assumed to be coupled to identical local phonon baths.
Thus, we consider a vibrational interaction Hamiltonian of the form:
\begin{equation} \label{eq:HIvib}
    \hat{H}_{I,vib} = \sum^{N}_{k=1}\hat{\sigma}_{k}^{z} \otimes \sum_{q} g_{q} \left(\hat{b}_{k, q} + \hat{b}_{k, q}^{\dag} \right)~,
\end{equation}
where $g_{q} \equiv g_{k,q}$ and $\hat{b}_{k, q}^{(\dag)}$ are the coupling strength and annihilation (creation) operator for the $q$th phonon mode with the $k$th dipole,  respectively.
Lastly, the optical and phonon environmental modes are governed by the following Hamiltonian:
\begin{equation} 
    \hat{H}_{f} = \sum_{p} \omega_{p} \hat{a}^{\dag}_{p} \hat{a}_{p} + \sum_{q, k} \tilde{\omega}_{q, k} \hat{b}^{\dag}_{q, k} \hat{b}_{q, k}~,
\end{equation}
where $\omega_{p}$ and $\tilde{\omega}_{q, k}$ are the frequencies of the $p$th photon modes and $q$th phonon modes at the $kth$ dipole.
The total Hamiltonian is then given by,
\begin{equation} \label{eq:Htot1}
     \hat{H} = \hat{H}_{S} + \hat{H}_{I,vib} +  \hat{H}_{I,opt} + \hat{H}_{f}~,
\end{equation}
describing the dynamics of chromophores as a collection of dipoles in interaction with each other, a shared optical bath, and individual local phonon baths.
We make the approximation that the interaction terms are weak and can therefore be used in a perturbative expansion wherein we truncate to second-order terms \cite{Breuer02Book, Schoffman20}.
The resulting Redfield quantum master equation describes the evolution of the density matrix $\rho$ of only the chromophore system:
\begin{equation}
\label{eq:QMEincohweak}
    \frac{\partial \rho}{\partial  t} = -i \left [\hat{H}, \rho \right] + \Gamma_{opt}\tilde{\mathcal{D}}_{opt}[\hat{\sigma}_{x}] + \Gamma_{vib}\tilde{\mathcal{D}}_{vib}[\hat{\sigma}_{z}]~ + \Gamma_{rad}\bar{\mathcal{D}}_{rad}[\hat{\sigma}^{-}_{t}]~,
\end{equation}
where $\Gamma_{vib}$ and $\Gamma_{opt}$ are the relaxation rates associated with the vibrational and optical baths, respectively.
As equations \eqref{eq:HIvib} and \eqref{eq:HIopt} take the form $\hat{H}=\hat{A}\otimes\hat{B}$ then the terms $\tilde{\mathcal{D}}_{opt}$ and $\tilde{\mathcal{D}}_{vib}$ are the non-secular Redfield dissipators for the photon and phonon fields, taking the form:
\begin{multline}
    \tilde{\mathcal{D}} = \sum_{n,m} \Gamma_{nm} \left(\omega_{m}\right) \left(A_{m}\left( \omega_{m}\right) \rho(t) A^{\dag}_{n}\left( \omega_{n}\right) \right. \\
    \left. -  A^{\dag}_{n}\left( \omega_{n}\right) A_{m}\left( \omega_{m}\right) \rho(t) + h.c. \right)~,
\end{multline}
where $A_{n}$ are the system operators associated with the dissipative process and the prefactors $\Gamma_{nm}(\omega)$ are environment correlation functions which depend on the relevant microscopic interactions with each environment \cite{Breuer02Book}.
As each vibrational environment is considered to be identical each correlation function is of the form:
\begin{equation}
    \Gamma_{nm}(\omega) = \frac{1}{2}\gamma_{nm}(\omega) + i S_{nm}(\omega).~
\end{equation}
We choose flat spectral density in order to construct the following rates for the vibrational dissipators;
\begin{equation} \label{eq:gamvib}
    \gamma^{vib}_{nm}\left(\omega\right) = \kappa^{vib}_{nm} \left(1 +  n_{vib}(\omega)\right),~
\end{equation}
whereas for the optical dissipators;
\begin{equation} \label{eq:gamopt}
    \gamma^{opt}_{nm}\left(\omega\right) = \kappa^{opt}_{nm} \left(1 +  n_{opt}(\omega)\right),~
\end{equation}
where $\kappa^{vib}_{nm}$ is a phenomenologically motivated prefactor chosen to reflect the timescale of the phonon rates and $\kappa^{opt}_{nm}$ is the spontaneous emission rate associated with the transition between the $n$th and $m$th energy eigenstates defined by the equation;
\begin{equation} \label{eq:kopt}
    \kappa^{opt}_{nm} = \frac{\omega_{nm}^{3} \textbf{d}_{n}\cdot \textbf{d}_{m}}{3 \pi \epsilon_{0} \hbar c^{3}}.~
\end{equation}
The functions $n_{vib}(\omega)$ and $n_{opt}(\omega)$ are the Bose-Einstein factors associated with the vibrational and optical baths respectively, each at the appropriate temperature of the associated environment. 
The final term of Eq.~(\ref{eq:QMEincohweak}), $\bar{\mathcal{D}}_{rad}[\hat{L}]$, is a Lindblad dissipator term describing radiative recombination, where $\Gamma_{rad}$ is the rate of this process, and takes the standard form:
\begin{equation}
    \bar{\mathcal{D}}(\rho)=\sum_{k}\hat{\sigma}^{-}_{k}\rho\hat{\sigma}^{+}_{k} - \frac{1}{2}\{\hat{\sigma}^{+}_{k}\hat{\sigma}^{-}_{k}, \rho\}~,
\end{equation}
for the lowering (raising) operators $\hat{\sigma}^{-}_{k}$ ($\hat{\sigma}^{+}_{k}$) acting upon  the $k$th site.
We use Equation \ref{eq:QMEincohweak} to resolve the system dynamics, in contrast to our previous work in Ref.~\cite{Schoffman20} which analysed equilibrium properties for the same type of model applied to the IsIA complex. This allows us to here investigate dynamic energy transport processes emerging from a physically-motivated model following the absorption of a photon.

It is important to highlight that the model presented in this work does not explore the limiting case a strong interaction between chromophores and the surrounding vibrational environment.
One could resolve this by moving to the polaron frame, which leads to an effective rescaling of the transition frequencies and hopping terms the bare Hamiltonian \cite{Nazir16, Brown19}.
However, the complexity of phycobilisome complexes and the number of sites required for a meaningful associated transport model makes this computationally demanding.
Furthermore, such a polaron approach would require more detailed knowledge of the particular spectral density present in PBS.
For these reasons, we employ the presented weak coupling model, which is sufficient to explore the basic nature of exciton transport in certain physical regimes whilst still providing a framework upon which more complex approaches can be built.

\subsection{Structural model} \label{sec:smod}

\begin{figure}[H]
    \centering
    \includegraphics[trim=0 0 0 0,clip, width=12cm]{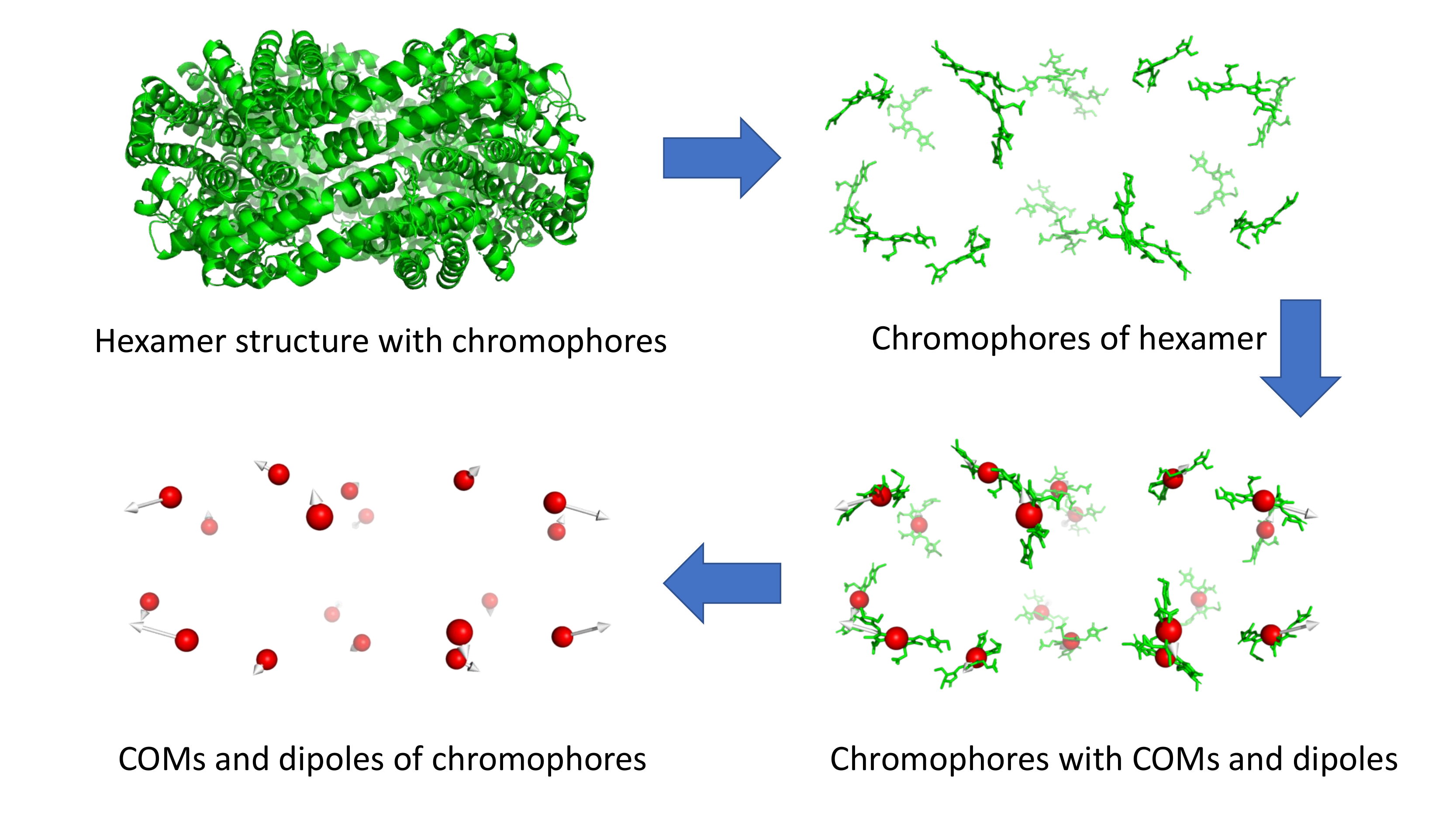}
    \caption{Schematic showing the generation of models to simulate energy distribution in the PBS antenna system. We started out with the \textit{S. elongatus} hexamer (PDB 4H0M), since the hexamer is the base unit of the antenna system. We then removed the scaffold proteins and left only the chromophores of the hexamer while maintaining their exact positions, as energy is transferred between the chromophores of the system. We aligned chromophore templates generated by Gaussian 09 to the hexamer chromophores according to type to determine the centers of mass and dipoles for the entire hexamer. Finally, we removed the chromophore structures and left only the centers of mass (COM) and dipole moment vectors (white arrows) for input into the simulation.}
    \label{fig:mod_diagram}
\end{figure}

From an extensive analysis of available structurally-resolved red algae PBS, it is clear that the hexamer structures throughout the system are uniform and the stacking of hexamers into rods is repetitive. This means that within an individual hexamer or rod, there, is little variation in the spacing and orientation between chromophores. However, the positioning of rods relative to each other compounds the variability of the structure, affecting spacing and orientation of inter-rod chromophores. Therefore, in this research we focused on the transfer of energy between  hexamers laterally, rather than within single rods or hexamers. For our models, we used the solved\textit{ Synechococcus elongatus }hexamer structure. We chose this hexamer, rather than one of the red algae rod hexamers, because of the simpler chemical properties of its chromophores and the smaller number of chromophores per hexamer. Due to the computational power needed for our simulation, we needed to simplify the representation of these complex molecules in our models.
The dipoles and centers of mass for each of the three chromophore types were calculated using Gaussian 09 \cite{g09}.
This data was used with the model discussed in the preceding section of the simulation to construct a physically-motivated description of system dynamics (see Figure \ref{fig:mod_diagram}).

\begin{figure}[H]
    \centering
    \includegraphics[trim=0 70 0 50,clip, width=12cm]{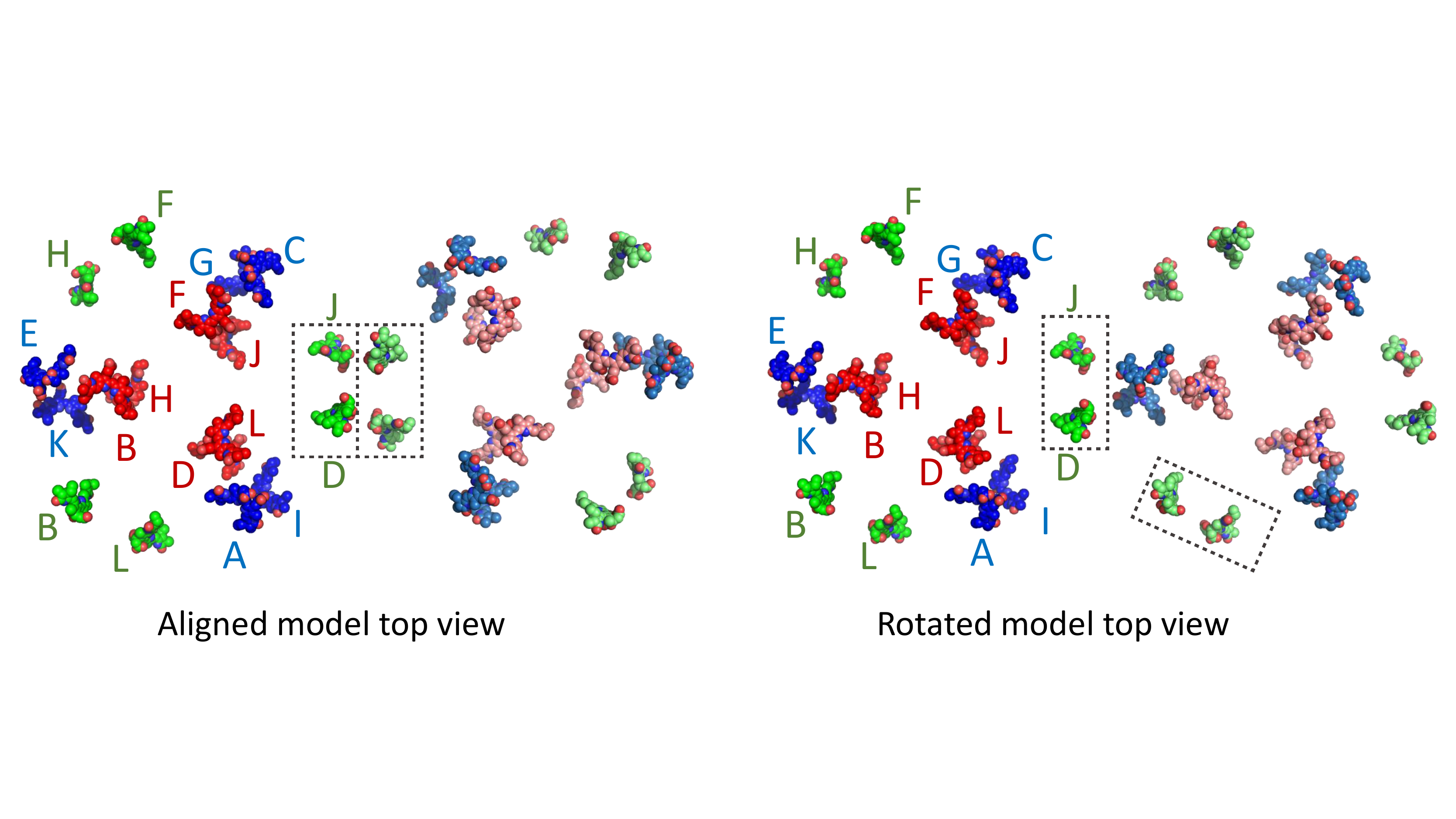}
    \caption{Models of the parallel hexamer pairs in both aligned (left) and rotated (right) orientations.
    Red; $\beta-84$ chromophores with a dipole vector magnitude of 9.836 Debye, blue; $\alpha-84$ chromophores with a dipole vector magnitude of 12.816 Debye, green; $\beta-155$ chromophores with a dipole vector magnitude of 10.709 Debye.
    Hexamer 1 chromophores are labeled and colored more brightly (on the left side of each model) and hexamer 2 chromophores are colored in a faded shade (on the right side of each model). In the aligned model, $\beta-155$ chromophores from both hexamers are present in the interface.
    In the rotated model, hexamer 2 is rotated 65\textsuperscript{$\circ$} so that its $\beta-155$ chromophores are staggered on either side of the interface (their positions are marked in dotted line boxes). }
    \label{aligned vs. rotated diagram}
\end{figure}

\begin{figure}[H]
    \centering
    \includegraphics[trim=0 70 0 70,clip, width=12cm]{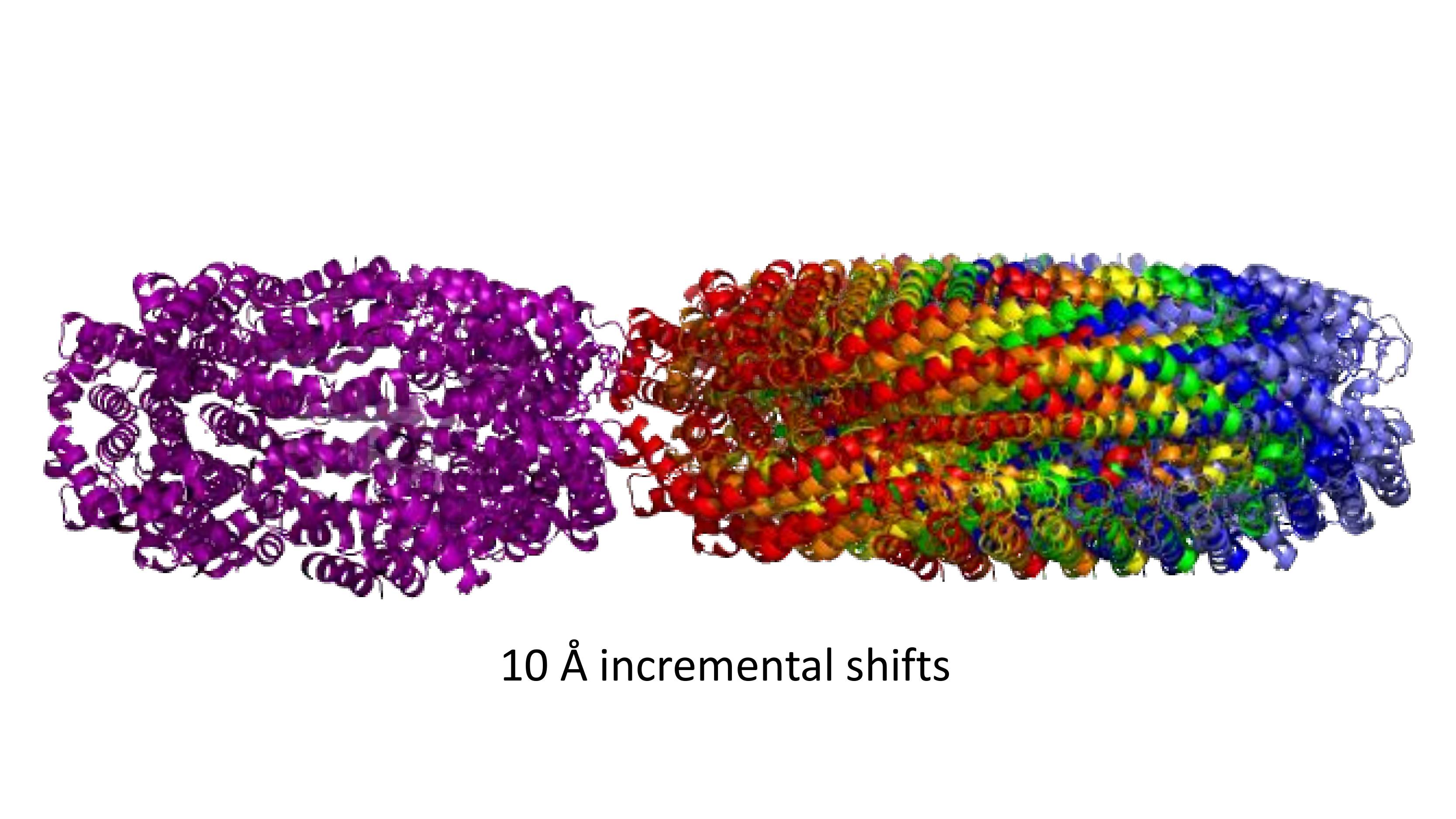}
    \caption{Hexamer pair model – Two copies of the S. elongatus hexamer, PDB 4H0M, were were placed in parallel next to each other as close as possible without overlapping. The hexamers were then moved apart at increments of 10 Å until reaching a final distance of 220 Å. Simulations were run on each incremental shift. Purple – hexamer 1; rainbow spectrum – hexamer 2 shifted at 10 Å increments away from hexamer 1. This Figure shows only the original position of hexamer 2 (red) and the first five shifts that reached a distance of 50 Å. }
    \label{10A shifts}
\end{figure}

To investigate the effect of distance and orientation on lateral energy transfer, we placed two parallel copies of the S. elongatus hexamer model as close together as possible without overlapping. For the aligned model, $\beta$-155 chromophores from both hexamers were positioned in the inter-hexamer interface. The rotated model was created by rotating the second hexamer of the aligned model 65\textsuperscript{$\circ$} (Figure \ref{aligned vs. rotated diagram}). We ran the simulation for both models, shifting the parallel hexamers further apart at 10~\AA{}  increments until reaching a final distance of 220~\AA{} between hexamers (Figure \ref{10A shifts}). 

\begin{figure}[H]
    \centering
    \includegraphics[trim=0 100 0 100,clip, width=12cm]{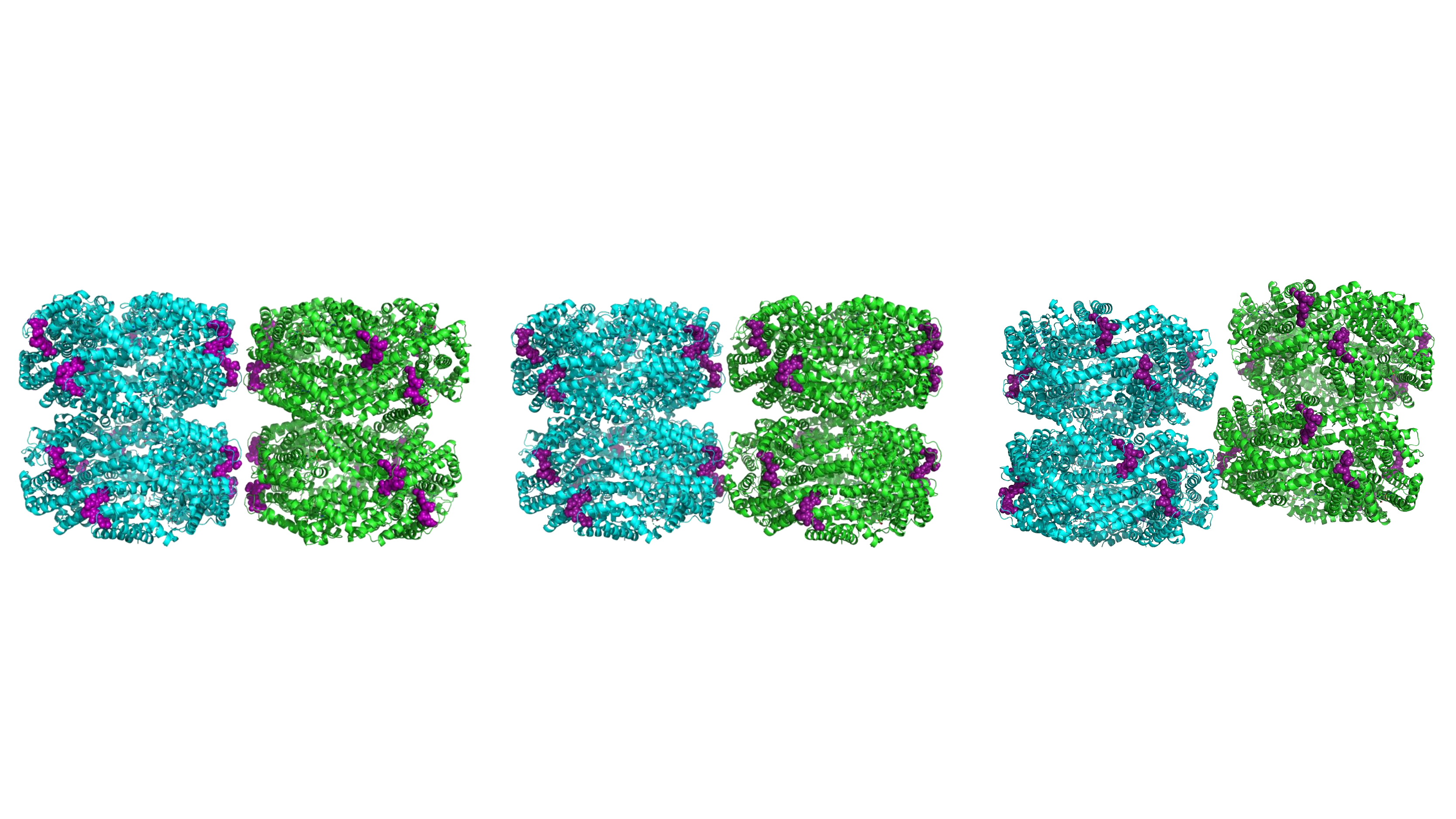}
    \caption{Rod pair models. Left - aligned rods, middle - rotated rods, right - real-world rods. Rod 1 for each pair is colored in cyan, rod 2 is colored in green. The $\beta-$155 chromophores in each model are colored in purple and shown as spheres. The rest of the system is shown in cartoon representation.}
    \label{rods diagram}
\end{figure}

The rod models were produced by adding a second hexamer below each of the original hexamers of the aligned and rotated hexamer models (see Figure \ref{rods diagram}). The purpose of this was to determine whether lateral energy transfer would be maintained in the presence of a vertical transfer option. The hexamers were stacked according to the stacking patterns found in the red algae \textit{P. purpureum} PBS, one of the only two publicly available sources for rod structures. A third rod model was created by aligning the \textit{S. elongatus} hexamer model to two hexamers of two neighboring rods of the \textit{P. purpureum} PBS. This was done to represent a real-world spatial positioning of rod pairs relative to each other, in contrast to the very artificial, parallel setups for the other models used in this study.
We used the artificial models to define control parameters for energy transfer in an intermediate coupled system, then evaluated the implementation of these parameters in the real-world structure. 

\section{Results} \label{sec:res}

\subsection{Control Parameters for Energy transfer Between Hexamers}

To evaluate the effect of distance and orientation on energy transfer, models were constructed to simulate the lateral transfer of energy between hexamer pairs.
For each simulation, energy transfer was initiated by exciting one of the 18 chromophores in hexamer 1. Under physiological light conditions the photon flux density is such that double excitation of a single PBS system are not expected and we therefore consider only the single excitation manifold \cite{Pollock13, Nazir09}.
The subsequent distribution of energy in both hexamer 1 and hexamer 2 is calculated across 100 picoseconds following initial excitation.
For these simulations we set the following phenomenologically-motivated rates: indicative phonon relaxation rate $\kappa_{vib}$, $1.0$ \si{ps}\textsuperscript{$-1$)}, and non-radiative recombination rate $\Gamma_{rad}$, $0.2$, \si{ns}\textsuperscript{$-1$} \cite{Curutchet17, Brown19, Schoffman20}.
The latter rate is chosen such that it competes with the optical rates, on the order of nanoseconds\textsuperscript{$-1$} and calculated using $\kappa_{opt}$, whereas the former is set such that the hierarchy of rates is appropriate, with vibrational rates being $10^{3}$ faster than the optical rates.

Pymol \cite{PyMOL} movies were used to visualize the energy distribution over time. These movies are provided in supplementary material. Movie1 and Movie2 show the energy distribution upon initially exciting an  $\alpha-$84 chromophore in the aligned and rotated models, respectively. As can be seen in the movies, most of the energy of the system is distributed between all the other $\alpha-$84 chromophores in both aligned and rotated models, although slightly less is laterally transferred to the $\alpha-$84 chromophores in hexamer 2, as compared to those in hexamer 1.
There is less but still significant energy transfer to the $\beta-$84 chromophores in hexamer 1, and a very small amount of transfer to the $\beta-$84 chromophores in hexamer 2. There was no significant distribution of energy to any of the $\beta-$155 chromophores.

Movie3 and Movie4 in the supplementary section show examples of energy distribution upon initially exciting a $\beta-$84 chromophore in the aligned and rotated models, respectively. Here we see the opposite effect as we saw after initially exciting the $\alpha-$84 chromophores. Most of the energy in the system is distributed between the $\beta-$84 chromophores, and to a lesser extent to the $\alpha-$84 chromophores. Here also, there is no significant distribution to the $\beta-$155 chromophores. 

It is therefore not surprising that following initial excitation of each of the $\beta-$155 chromophores (see Movie5 and Movie6 in the supplementary section for the aligned and rotated models, respectively), energy distribution is isolated to the other $\beta-$155 chromophores of the system for both aligned and rotated models. Significantly, when an interface $\beta-$155 chromophores is initially excited in the aligned model, there is localization of energy. That is, despite the relatively weak coupling inherent to the system, almost all the energy remains trapped between four $\beta-$155 chromophores.   

\begin{figure}[H]
    \centering
    \includegraphics[trim=100 0 100 0,clip, width=12cm]{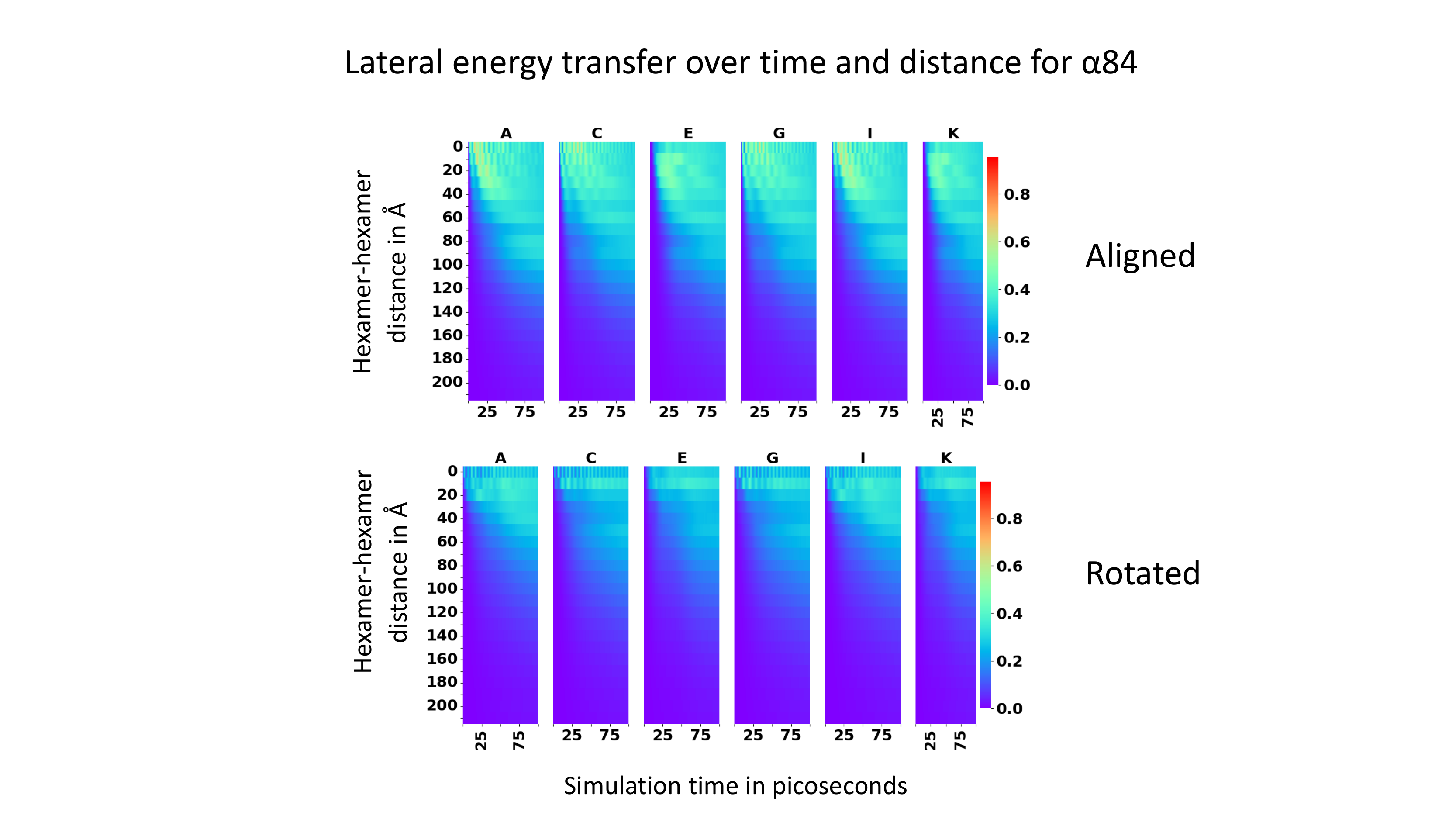}
    \caption{Heatmap showing the lateral energy transfer from hexamer 1 to hexamer 2 over time (x-axis) and distance (y-axis) following initial excitation of each $\alpha-84$ chromophore in hexamer 1. The color bar measures the amount of energy transferred out of the total energy of the system. 1.0 signifies all of the total energy, and 0.0 signifies none of the total energy. Thus, dark blue indicates low energy transfer, while red indicates high energy transfer.}
    \label{hexamer alpha84 heatmaps}
\end{figure}

To break down what we are seeing in the simulation further, the heatmaps in Figure \ref{hexamer alpha84 heatmaps} show the lateral energy transfer from hexamer 1 to hexamer 2 over time and distance following initial excitation of the $\alpha-$84 chromophores on hexamer 1.
At close proximity, we observe rapid oscillations of lateral transfer between the hexamers (chains A, C, G, and I in the aligned and rotated models).
Over larger distances, both oscillation frequency and overall lateral transfer declines.
At close proximity in the aligned model, more energy is transferred to hexamer 2 in the first 50 picoseconds of the simulation, and less is transferred to hexamer 2 as time continued, particularly in chains A, C, G, and I. At larger distances, most energy is transferred to hexamer 2 in the second half of the simulation.
Overall, the patterns of lateral energy transfer are similar in both the aligned and rotated models, although slightly more energy transferred laterally in the aligned model.
\begin{figure}[H]
    \centering
    \includegraphics[trim=100 0 100 0,clip, width=12cm]{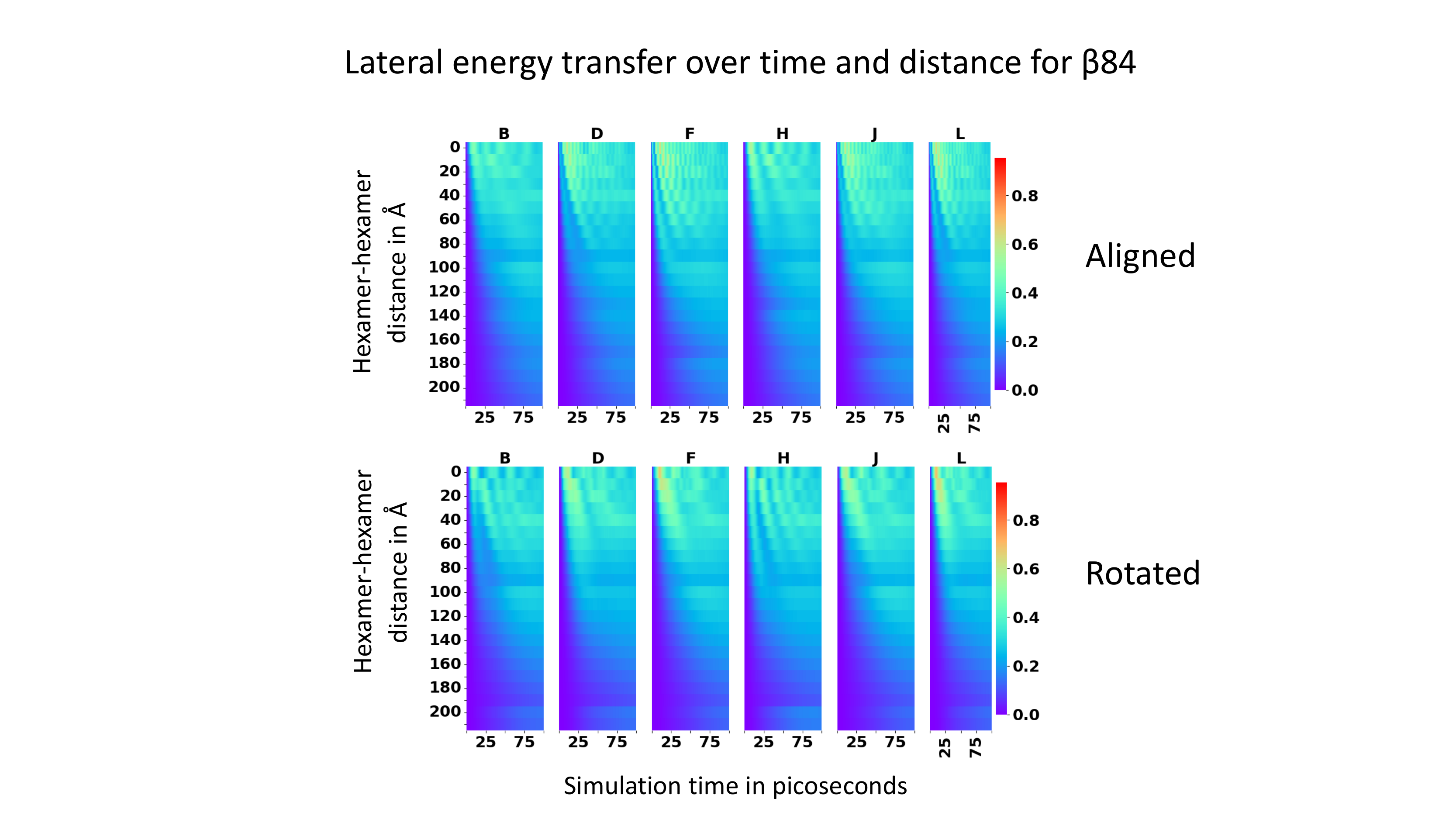}
    \caption{Heatmap showing the lateral energy transfer from hexamer 1 to hexamer 2 over time (x-axis) and distance (y-axis) following initial excitation of each $\beta-84$ chromophore in hexamer 1. The color bar measures the amount of energy transferred out of the total energy of the system. 1.0 signifies all of the total energy, and 0.0 signifies none of the total energy. Thus, dark blue indicates low energy transfer, while red indicates high energy transfer.
    \label{hexamer beta84 heatmap}
}
\end{figure}

Following initial excitation of the $\beta-$84 chromophores on hexamer 1 (Figure \ref{hexamer beta84 heatmap}), lateral energy transfer is also characterized by rapid oscillation when hexamer 2 was in close proximity to hexamer 1, especially in chains D, F, J, and L of the aligned model. However, this oscillation is not nearly as rapid as what was seen for the $\alpha-$84 chromophores.
The interior position of the $\beta-$84 chromophores in the core of the hexamer, in contrast to the more exterior position of the $\alpha-$84 chromophores, could explain this reduced short-range oscillation \ref{aligned vs. rotated diagram}.
However, there is still very strong long-range lateral energy transfer to hexamer 2 at much greater distances following initial excitation of the $\beta-$84 chromophores, compared to what was simulated for the $\alpha-$84 chromophores. 

\begin{figure}[H]
    \centering
    \includegraphics[trim=100 0 100 0,clip, width=12cm]{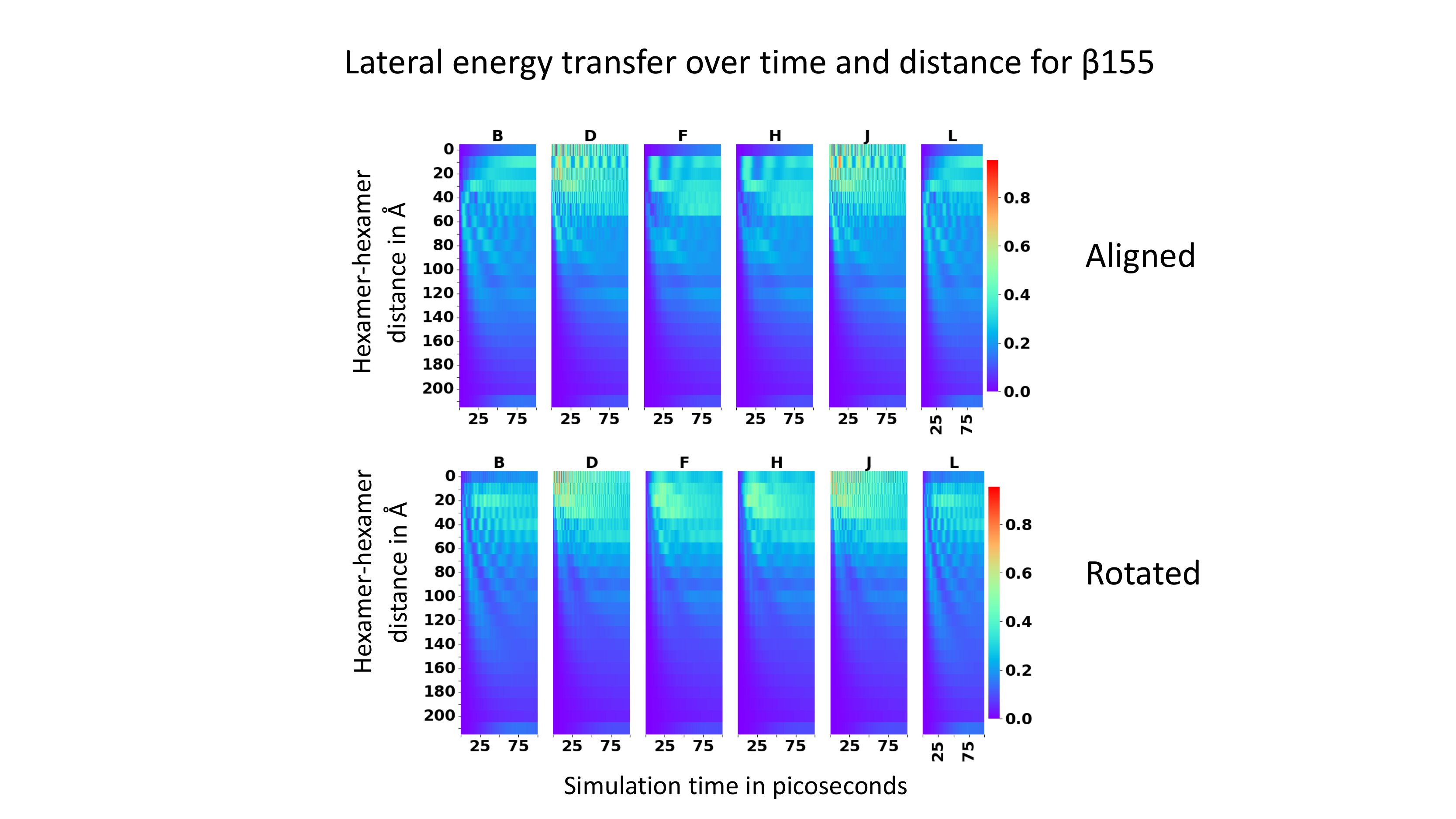}
    \caption{Heatmap showing the lateral energy transfer from hexamer 1 to hexamer 2 over time (x-axis) and distance (y-axis) following initial excitation of each $\beta-155$ chromophore in hexamer 1. The color bar measures the amount of energy transferred out of the total energy of the system. 1.0 signifies all of the total energy, and 0.0 signifies none of the total energy. Thus, dark blue indicates low energy transfer, while red indicates high energy transfer.}
    \label{hexamer beta155 heatmap}
\end{figure}

When either of the two hexamer 1 $\beta-$155   interface chromophores are initially excited (Figure \ref{hexamer beta155 heatmap}), extremely rapid energy oscillation and energy transfer occurs. This can be seen in chains D and J of the aligned and rotated models. 
However, when any of the other four chromophores on hexamer 1 (chains B, F, H, and L) are initially excited, there was very little energy transfer at distance 0, although energy transfer does increase between 10 \AA{} and about 40 \AA{}. This is the case in both aligned and rotated models. 

After summing the total lateral energy transfer for each distance, we see that following initial excitation of many of the $\alpha-$84 and $\beta-$155 chromophores, the maximum total energy transfer does not occur when the two hexamers are at distance 0 (see Figure S1). Rather, the optimal distance for lateral energy transfer can be as much as 30 \AA{} from the original position. This shows that while there is a general correlation between distance and lateral energy transfer (the closer the hexamers, the higher the lateral energy transfer), lateral energy transfer is somewhat diminished when the interface chromophores are too close. Further explanation can be found in the supplementary section.  

Overall, we can see from our results that for both aligned and rotated hexamer models energy transfer is not sensitive to moderate changes in orientation, provided the positioning of the chromophores in the interface does not result in strong energy localization. There is a small degree of difference in the optimal distances for lateral energy transfer between the aligned and rotated models. There is slightly more energy transfer in the aligned model following initial excitation of an $\alpha$-84 chromophore, but in general energy transfer seems to be very robust and flexible, tolerating a fair amount of variation with relatively constant transfer patterns. Distance obviously controls the rate of transfer, but the correlation is not trivial. Very short hexamer-hexamer distances can be detrimental to transfer efficiency due to the trapping of the energy in local states between chromophore pairs in the interface. 

\subsection{Control Parameters for Energy transfer Between Rods}

While simulating the energy transfer in hexamer pairs gave us a good start in understanding the control parameters of our system, most of the PBS is made up of rods along which energy is transferred from the periphery to the core of the structure. We therefore placed an additional hexamer below each of the original hexamers of our models to add the possibility of vertical transfer to our model (see \ref{rods diagram}).
This procedure created simple rod models in both the aligned and rotated orientations. 

We also created a more ``real-world" model by locating two neighboring rods in the publicly-available EM resolved \textit{P. purpureum} structure (PDB 6KGX)  and aligning our cyanobacteria hexamer to two red algae hexamers in each rod. We chose to create a homology based model instead of using the \textit{P. purpureum} hexamer itself, as the rods of the \textit{P. purpureum} antenna system bind phycoerythrobilin or phycourobilin chromophores instead of phycocyanobilin chromophores.  Moreover, the \textit{P. purpureum} hexamers contain more chromophores than \textit{S. elongatus} hexamers, which would have increased the complexity of the simulation considerably. ``Real-world" rods are not parallel as in our aligned and rotated models, rather they have a staggered orientation (see Figure \ref{rods diagram}). 
As was done with the two hexamer model simulations, each of the 18 chromophores in hexamer 1 were initially excited, and energy distribution throughout the rod pairs was simulated.
Here also, energy distribution movies were used to visualize the energy distribution in the rod systems.

Movie7, Movie8, and Movie9 in the supplementary section display the distribution in the system after initially exciting an $\alpha$-84 chromophore in the aligned, rotated, and real-world models, Movie10, Movie11, and Movie12 show distribution after initially exciting a $\beta-84$ chromophore in the aligned, rotated, and real-world models, respectively, and Movie13, Movie14, and Movie15 represent the distribution after initially exciting a $\beta-$155 chromophore in the aligned, rotated, and real-world models, respectively. For the aligned and rotated models, these movies clearly show lateral energy transfer patterns that are very similar to those of their hexamer pair counterparts, despite the presence of a vertical transfer option.

\begin{figure}[H]
    \centering
    \includegraphics[trim=0 80 0 0,clip, width=12cm]{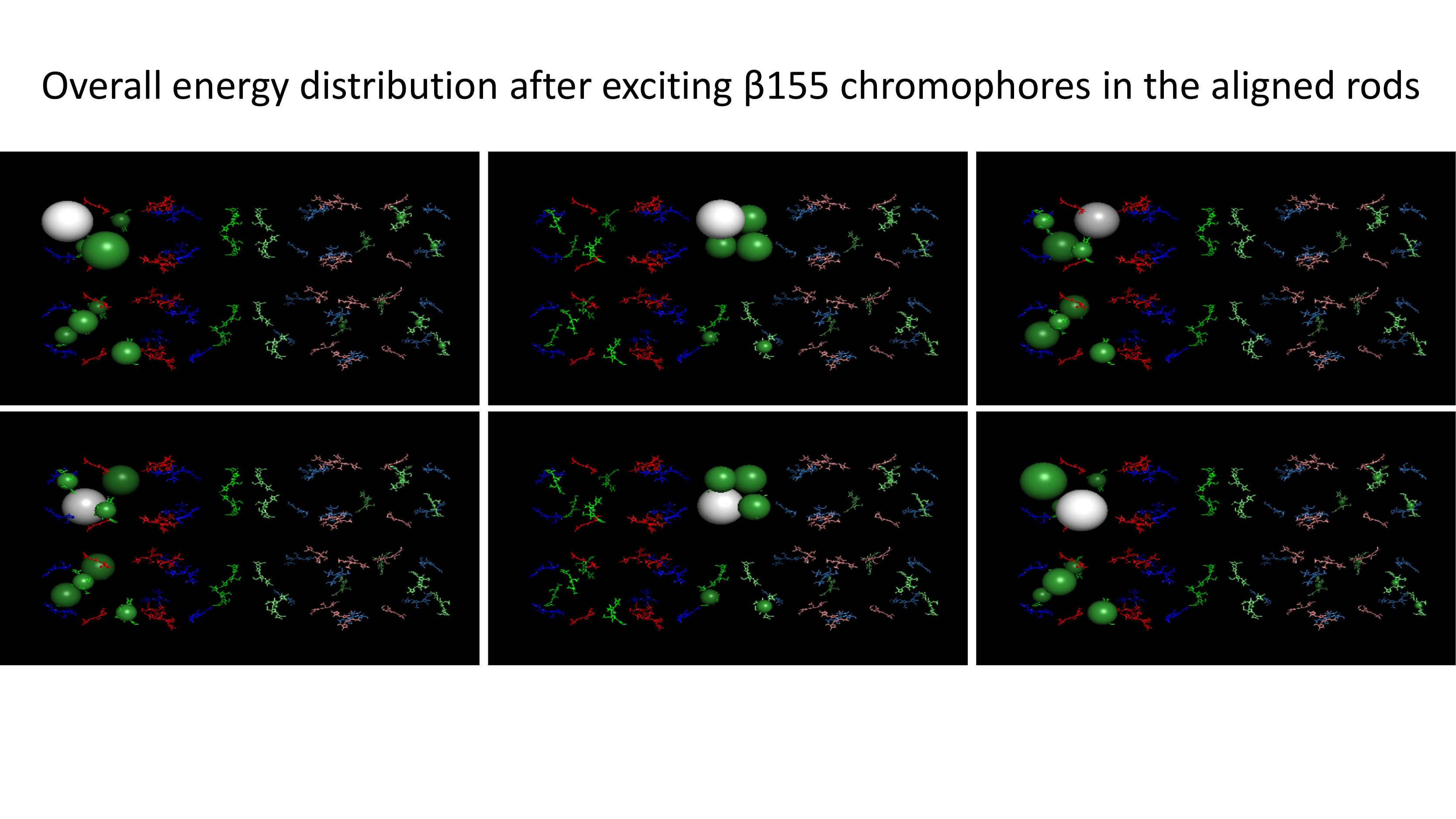}
    \caption{Pymol visualization of energy distribution after initial excitation of all six $\beta-155$ chromophores in the aligned rod model. Chromophores are shown as sticks and colored according to type ($\alpha-84$ chromophores in blue, $\beta-84$ chromophores in red, and $\beta-155$ chromophores in green). The amount of energy distributed to an individual chromophore is represented by a sphere, the size of which indicates the highest amount of energy transferred to that specific chromophore at any point during the 100-picosecond simulation. Sphere colors match chromophore type, except for the white sphere, which marks the initially excited chromophore. See the methods section for more details. 
}
    \label{static aligned beta155}
\end{figure}

To provide an overview of the energy distribution in all the rod systems after initial excitation of all 18 chromophores in hexamer 1, we created a static visualization similar to the movies described above. One difference in the static visualization, however, is that the size of the sphere indicates the highest amount of energy transferred to a given chromophore at any point during the 100-picosecond simulation.

\begin{figure}[H]
    \centering
    \includegraphics[trim=0 80 0 0,clip, width=12cm]{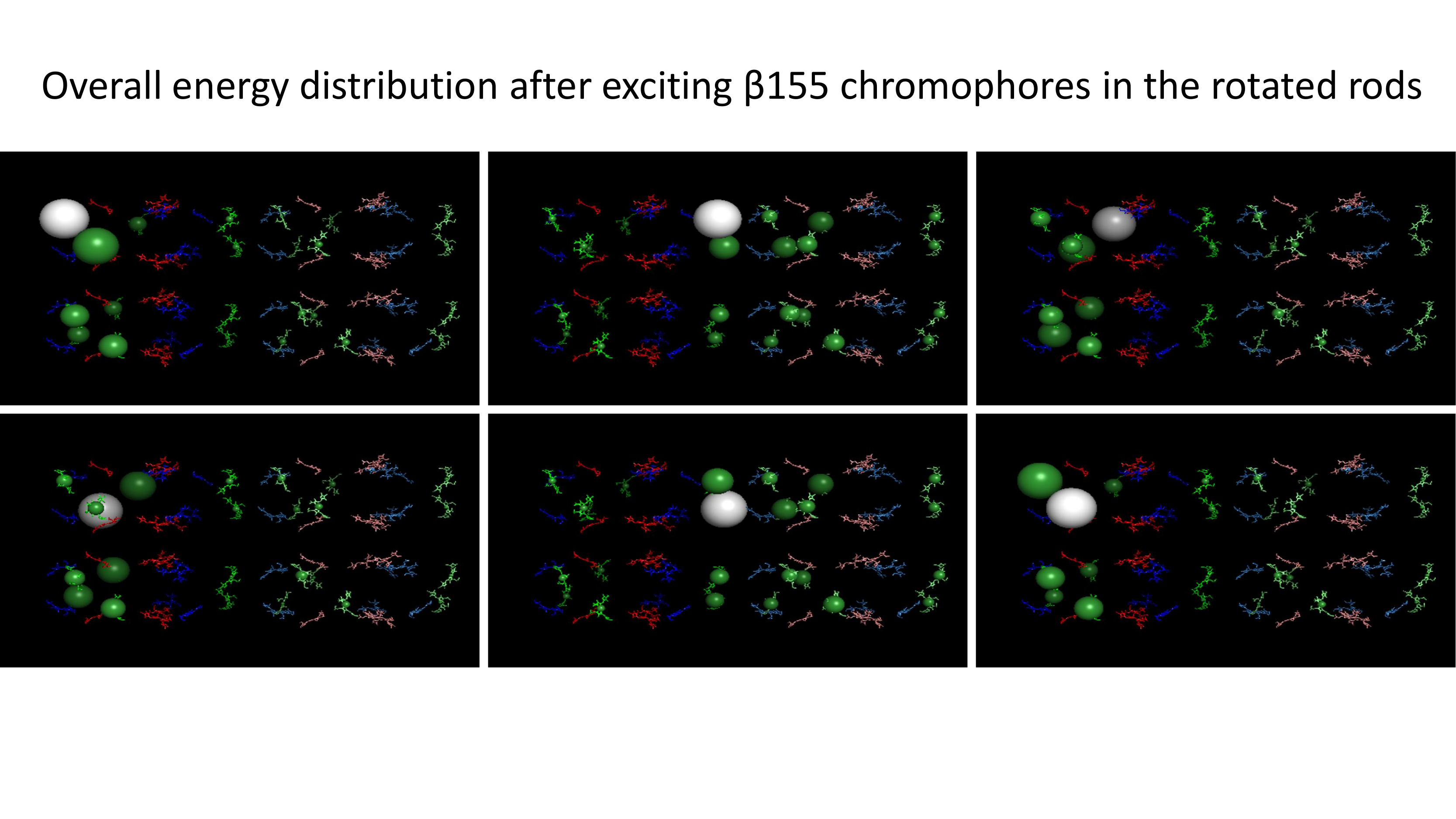}
    \caption{Pymol visualization of energy distribution after initial excitation of all six $\beta-155$ chromophores in the rotated rod model. Chromophores are shown as sticks and colored according to type ($\alpha-84$ chromophores in blue, $\beta-84$ chromophores in red, and $\beta-155$ chromophores in green). The amount of energy distributed to an individual chromophore is represented by a sphere, the size of which indicates the highest amount of energy transferred to that specific chromophore at any point during the 100-picosecond simulation. Sphere colors match chromophore type, except for the white sphere, which marks the initially excited chromophore. See the methods section for more details.}
    \label{static rotated beta155}
\end{figure}

\begin{figure}[H]
    \centering
    \includegraphics[trim=0 80 0 0,clip, width=12cm]{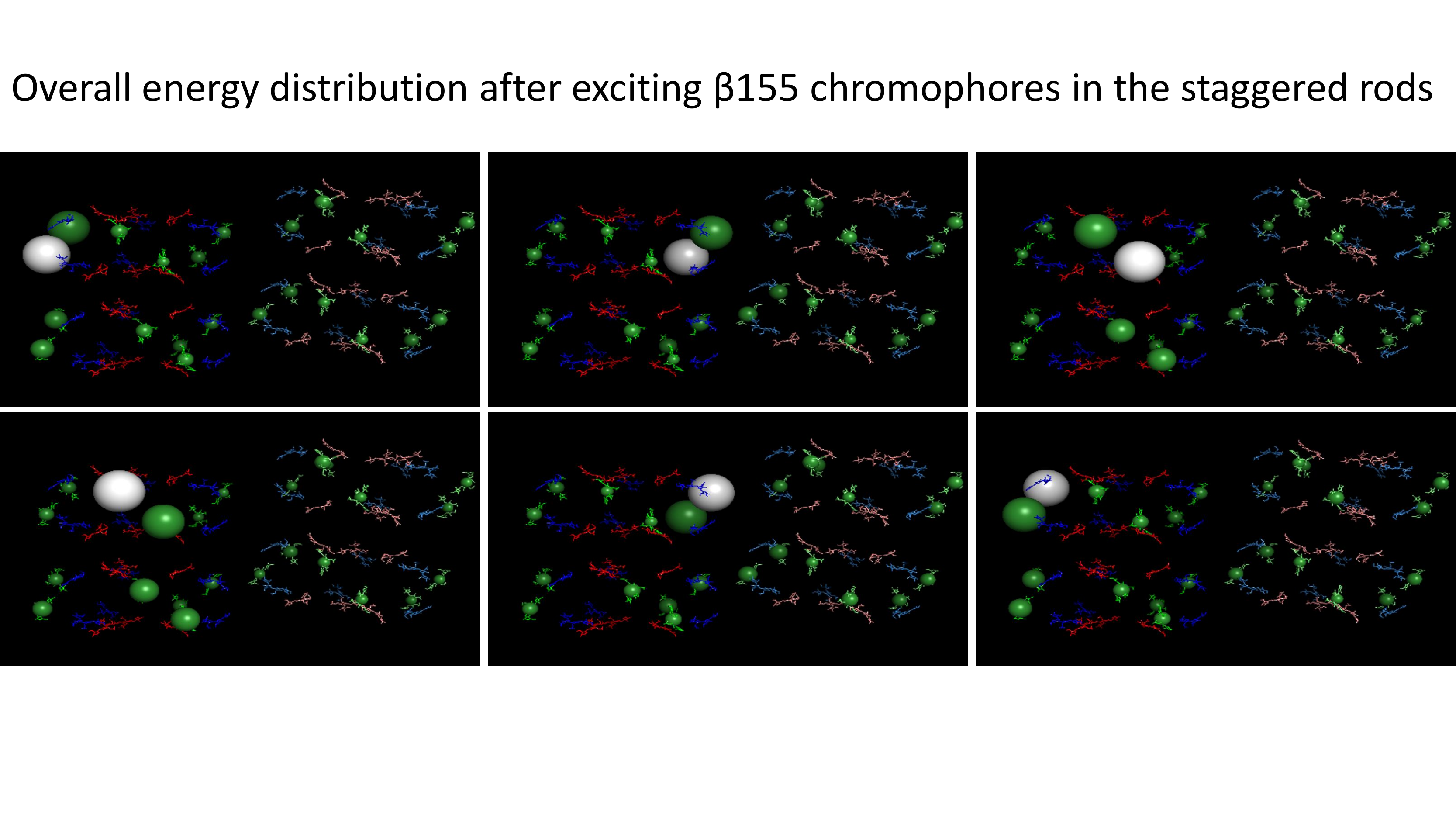}
    \caption{Pymol visualization of energy distribution after initial excitation of all six $\beta-155$ chromophores in the real-world rod model. Chromophores are shown as sticks and colored according to type ($\alpha-84$ chromophores in blue, $\beta-84$ chromophores in red, and $\beta-155$ chromophores in green). The amount of energy distributed to an individual chromophore is represented by a sphere, the size of which indicates the highest amount of energy transferred to that specific chromophore at any point during the 100-picosecond simulation. Sphere colors match chromophore type, except for the white sphere, which marks the initially excited chromophore. See the methods section for more details.
}
    \label{static staggered beta155}
\end{figure}

The most apparent differences in transfer patterns between models can be seen following initial excitation of all six of the $\beta-$155 chromophores (Figures \ref{static aligned beta155}, \ref{static rotated beta155}, and \ref{static staggered beta155}). In the aligned models, the energy distribution remains in the area of the initially excited chromophore, and significant lateral transfer only occurs when the initially excited chromophore is located in the inter-rod interface.
When this happens, energy is trapped between a few very local $\beta-$155 chromophores.
These patterns play out in the rotated model as well, but the energy is trapped in the interface to a lesser degree and spreads out more into rod 2, most likely due to the larger distance between the $\beta-$155 chromophores in the interface.
Significantly, in the real-world model, there is a much more even distribution between all the $\beta-$155 chromophores in both rods.
This could be because the staggered orientation causes  most of the $\beta-$155 chromophores in the interface to move slightly further apart, although there are still several interface $\beta-$155 chromophores that are as close to each other in the staggered model as they are in the rotated model. 

The simulations of short rod structures provide additional insight into energy transfer patterns. Importantly, both lateral and horizontal transfer are mostly unaffected by changes in the hexamer unit orientation, provided close localization of pigment pairs is avoided. The staggered, angled positioning of the real-world rods seems specifically configured to avoid such localizations.  

\section{Conclusions} \label{sec:concl}

At first glance, massively pigmented intermediately-coupled PBS systems do not seem to fit as effective light harvesting systems. However, the simulations presented in this paper provide additional insights that demonstrate certain advantages of such structures over tightly-coupled pigment systems. As long as closely localized hexamer pairs are avoided, as in the “real-world” model, energy transfer is robust.  Energy is distributed relatively evenly both horizontally and vertically to the rod axis, regardless of the type of chromophore excited and its location in the structure.  A single PBS system is large enough to provide excition energy to more than one photosystem \cite{Ueno16}. Therefore, the ability to evenly distribute energy between rods and over the entire PBS can play an important biological role.  
Furthermore, energy transfer is relatively immune to the effects of distances or rotations, within the range of intermediate coupling distances. This robustness is critical in the context of large biological structures operating at room temperature in living cells.  Considering the erratic nature of the light field in marine environments \cite{Kolodny22}, it is easy to see how the robustness and the ability to distribute energy over the large volume of the PBS structure can convey advantages to marine photosynthesis. 

\section{Methods}

\subsection{Static Pymol visualization}

The static Pymol visualization was used to create Figures \ref{static aligned beta155}, \ref{static rotated beta155}, and \ref{static staggered beta155}.
Each chromophore was shown as sticks and colored according to type ($\alpha-$84 chromophores in blue, $\beta-$84 chromophores in red, and $\beta-$155 chromophores in green), except for the white sphere, which marks the chromophore that was initially excited during the simulation.

The amount of energy distributed to each individual chromophore was represented by a sphere located at its center of mass coordinates.
Ranges for sphere size were given according to the highest proportion of energy that transferred to a given chromophore at any point during the 100-picosecond simulation (see Table \ref{tab:sphere_sizes}). If the highest amount of energy for a given was less than .002, no sphere was generated to represent the chromophore.

\begin{table}[h!]
\centering
\begin{tabular}{||c | c ||}
 \hline
 Proportion of energy & Sphere size in Pymol \\
 \hline \hline
 $\geq0.9$ & 12  \\
 \hline
 $0.8\rightarrow0.9$ & 11  \\
 \hline
 $0.7\rightarrow0.8$ & 10  \\
 \hline
 $0.6\rightarrow0.7$ & 9  \\
 \hline
 $0.5\rightarrow0.6$ & 8  \\
 \hline
 $0.4\rightarrow0.5$ & 7  \\
 \hline
 $0.3\rightarrow0.4$ & 6  \\
 \hline
 $0.2\rightarrow0.3$ & 5  \\
 \hline
 $0.1\rightarrow0.2$ & 4  \\
 \hline
 $0.05\rightarrow0.1$ & 3  \\
 \hline
 $0.025\rightarrow0.05$ & 2  \\
 \hline
 $0.002\rightarrow0.025$ & 1  \\
 \hline
 $0.0\rightarrow0.002$ & 0  \\
 \hline
\end{tabular}
\caption{Chart showing the Pymol sphere size for the proportion of energy transferred to a chromophore.}
\label{tab:sphere_sizes}
\end{table}

\subsection{Dynamic Pymol visualization}

The dynamic Pymol visualization was used to create movies 1-15 in the supplemental section. The methodology for generating the dynamic visualization using Pymol was similar to that of the static visualization, except that instead of using a sphere to represent the largest amount of energy distributed to each chromophore over the 100 ps simulation, the exact amount of energy transferred to each chromophore for every 0.1 ps timepoint was represented by a sphere in a separate video frame (see Table \ref{tab:sphere_sizes}). The videos have a total of 1000 frames showing energy transfer over the course of the 100 ps simulation.

\subsection{Transition dipole moment calculations}

The magnitude and orientation of the transition dipole moment of each chromophore in the system are used in the Equations \eqref{eq:forst} and \eqref{eq:kopt} to calculate the interaction between dipoles.
These calculations are necessary for our simulation of the overall energy flow of the PBS antenna system.
The transition dipole moments for the three protonated phycocyanobilin conformers were obtained using quantum chemistry calculations.
Heavy-atom coordinates were used from the Protein Data Bank \cite{Berman03} with explicit hydrogen atoms incorporated in a constrained geometry optimization.
Density functional theory calculations were carried out at the B3LYP level using the 6-31G(d) basis, with the five lowest electronic excited states computed via time-dependent methods.
All calculations used Gaussian 09 \cite{g09}.

\bibliographystyle{unsrt}
\bibliography{main}

\section*{Acknowledgments} 
NK and EJD were supported by the Israel Science Foundation grant 1182/19 and the Zelman Cowen Academic Initiatives.

NW and EMG acknowledge funding from the EPSRC grant no.~EP/T007214/1.

The authors acknowledge the University of Maryland supercomputing resources (http://hpcc.umd.edu) made available for conducting the research reported in this paper.

Special thanks to Prof. Leah Dodson for her contribution and kind support and to Noa Hazony for her teamwork and technical support. 
\end{document}